\documentclass[12pt]{article}
\usepackage{amssymb}
\usepackage{amsmath}
\usepackage{amsfonts}
\usepackage{epsfig}
\usepackage{anysize}
\marginsize{2cm}{2cm}{2.5cm}{3cm}
\date{}

\begin{document}

\title{\textbf{Emergent statistical-mechanical structure in the dynamics along the period-doubling route to chaos}}

\author{Alvaro Diaz-Ruelas\textsuperscript{1}, Alberto Robledo\textsuperscript{1,2}\\
\tiny 1. Instituto de F\'{i}sica, Universidad Nacional Aut\'{o}noma de M\'{e}xico \\
\tiny Apartado Postal 20-364, M\'{e}xico 01000 D.F., Mexico.\\
\tiny 2. Centro de Ciencias de la Complejidad, Universidad Nacional Aut\'{o}noma de M\'{e}xico\\
\tiny Apartado Postal 20-364, M\'{e}xico 01000 D.F., Mexico.\\
       }
        
\maketitle

\abstract{\footnotesize We consider both the dynamics within and towards the supercycle attractors along the period-doubling route to chaos to analyze the development of a statistical-mechanical structure. In this structure the partition function consists of the sum of the attractor position distances known as supercycle diameters and the associated thermodynamic potential measures the rate of approach of trajectories to the attractor. The configurational weights for finite $2^{N}$, and infinite $N \rightarrow \infty $, periods can be expressed as power laws or deformed exponentials. For finite period the structure is undeveloped in the sense that there is no true configurational degeneracy, but in the limit $N\rightarrow \infty $ this is realized together with the analog property of a Legendre transform linking entropies of two ensembles. We also study the partition functions for all $N$ and the action of the Central Limit Theorem via a binomial approximation.}     
\\   

\vspace*{6pt}

\noindent 
{\tiny PACS 5.45.Ac Low-dimensional chaos}\\
{\tiny PACS 05.20.Gg Classical ensemble theory}\\
{\tiny PACS 05.45.Df Fractals}\\

      %%%%%%%  SECTION 1.
\section{Introduction} 
\label{Sec1} %%%%%%%%%% 1. INTRODUCTION %%%%%%%%

For thermal systems formed by particles interacting via standard forces the
limit of validity of equilibrium statistical mechanics is, trivially,
non-equilibrium. Thermal systems constitute the normal realm of the Boltzmann-Gibbs 
(BG) formalism, but there are other types of systems for which it has been known
for some time that they accept a statistical-mechanical description of the
BG type. These are multifractals and chaotic nonlinear dynamical systems 
\cite{beck1}, among which one-dimensional unimodal iterated maps, represented by the
quadratic logistic map, are familiar model systems \cite{schuster1, hilborn1} that exhibit
such properties. The chaotic attractors
generated by this class of maps have ergodic and mixing properties and not
surprisingly they can be described by a thermodynamic formalism compatible
with BG statistics \cite{beck1}. But at the transition to chaos, the
period-doubling accumulation point, the so-called Feigenbaum point, these
two properties are lost and this suggests the possibility of exploring the 
limit of validity of the BG structure in a precise but simple enough setting. 

Recently a comprehensive description has been given \cite{robledo1, robledo2}
of the elaborate dynamics that takes place both inside and
towards the Feigenbaum attractor. Amongst several conclusions, these studies
established that the two types of dynamics are related to each other in a
statistical-mechanical way, \textit{i}.\textit{e}. the dynamics at the
attractor provides the `microscopic configurations' in a partition function
while the approach to the attractor is efficiently described by an entropy
obtained from it. As we show below, this property conforms to $q$
-deformations \cite{robledo1,robledo2,tsallis1,tsallis2},  of the ordinary exponential
weight of BG statistics. This novel statistical-mechanical feature arises
in relation to a multifractal attractor with vanishing Lyapunov exponent.
Here we explore in more detail this property with focus on how the
statistical-mechanical structure develops along the period-doubling
bifurcation cascade \cite{schuster1, hilborn1}, i.e. out of chaos.

Deformed exponentials appear in the studies of many physical systems. For instance, simulated velocity distributions of statistical-mechanical models resemble closely the so-called $q$-gaussian expression \cite{hu1, hu2}, suggesting the occurrence of generalized statistical-mechanical structures under non-equilibrium conditions. Here, as an effort to provide a firm basis to a wider content discussion, we chose to study a nontrivial archetypal system under ergodicity and mixing failure and precisely determine its properties independently of any method that assumes a statistical-mechanical formalism. After that, the results obtained can be analyzed in relation to generalized entropy expressions or properties derived from them.

%%%%%%%%%%%%%%%%%%%%%%%%%%%%%   SECTION 2.   %%%%%%%%%%%%
																				
\section{Brief recall of the dynamics within and towards the Feigenbaum attractor} \label{Sec2}   

%%%%%%%%%%% 2. Brief recall of the dynamics within and towards ... %%

The trajectories associated with the period-doubling route to chaos in
unimodal maps exhibit elaborate dynamical properties that follow concerted
patterns. At the period-doubling accumulation points, periodic attractors become multifractal before turning chaotic. At these points the Lyapunov exponent $\lambda$ vanishes as it
changes sign \cite{schuster1, hilborn1}. There are two sets of
properties associated with the attractors involved: those of the dynamics
inside the attractors and those of the dynamics towards the attractors.
These properties have been characterized in detail, the organization of
trajectories and also that of the sensitivity to initial conditions at the
Feigenbaum attractor are described in Ref. \cite{robledo1}, while the
features of the rate of approach of an ensemble of trajectories to this
attractor has been explained in Ref. \cite{robledo2}.

We recall some of the basic features of the bifurcation forks that form the
period-doubling cascade sequence in unimodal maps, often illustrated by the
logistic map $f_{\mu }(x)=1-\mu x^{2}$, $-1\leq x\leq 1$, $0\leq \mu \leq 2$ 
\cite{schuster1, hilborn1}. The knowledge of the dynamics towards a
particular family of periodic attractors, the so-called superstable
attractors \cite{schuster1, hilborn1}, facilitates the understanding
of the rate of approach of trajectories to the Feigenbaum attractor, located
at $\mu =\mu _{\infty }=1.401155189092...$, and highlights the source of the
discrete scale invariant property of this rate \cite{robledo2}. The family
of trajectories associated with these attractors - also called supercycles -
of periods $2^{N}$, $N=1,2,3,...$, are located along the bifurcation forks.
The positions (or phases) of the $2^{N}$- attractor are given by $x_{j}=f_{%
\overline{\mu }_{N}}^{(j)}(0)$, $j=1,2,\ldots ,2^{N}$. Associated with the $2^{N}$-attractor at $\mu =\overline{\mu }_{N}$ there is a $(2^{N}-1)$-repellor consisting of $2^{N}-1$ positions $y_{k}$, $k=0,1,2,\ldots ,2^{N}-1$. These positions are the
unstable solutions, $\left\vert df_{\overline{\mu }_{N}}^{(2^{n-1})}(y)/dy%
\right\vert > 1$, of $y=f_{\overline{\mu }_{N}}^{(2^{n-1})}(y)$, $%
n=1,2,\ldots ,N$. The first, $n=1$, originates at the initial
period-doubling bifurcation, the next two, $n=2$, start at the second
bifurcation, and so on, with the last group of $2^{N-1}$, $n=N$, setting out
from the $N-1$ bifurcation. The diameters $d_{N,m}$ are defined as 
$d_{N,m} \equiv x_{m} - f_{\overline{\mu }_{N}}^{(2^{N-1})}(x_{m})$ \cite{schuster1,hilborn1}.

Central to our understanding of the dynamical properties of unimodal maps is
the following in-depth property: Time evolution at $\mu _{\infty }$ from $%
\tau =0$ up to $\tau \rightarrow \infty $ traces the period-doubling cascade
progression from $\mu =0$ up to $\mu _{\infty }$. There is an underlying
quantitative relationship between the two developments. Specifically, the
trajectory inside the Feigenbaum attractor with initial condition $x_{0}=0$,
the $2^{\infty }$-supercycle orbit, takes positions $x_{\tau }$ such that
the distances between appropriate pairs of them reproduce the diameters $%
d_{N,m}$ defined from the supercycle orbits with $\overline{\mu }_{N}<\mu
_{\infty }$. See Fig. 1 in Ref. \cite{robledo2}. This property has
been basic in obtaining rigorous results for the sensitivity to initial
conditions for the Feigenbaum attractor \cite{robledo1}, and for the
dynamics of approach to this attractor \cite{robledo2}. Other families of
periodic attractors share most of the properties of supercycles.

The organization of the total set of trajectories as generated by all
possible initial conditions as they flow towards a period $2^{N}$ attractor
has been determined in detail \cite{robledo2, robledo3}. It was found
that the paths taken by the full set of trajectories in their way to the
supercycle attractors (or to their complementary repellors) are
exceptionally structured. The dynamics associated to families of
trajectories always displays a characteristically concerted order in which
positions are visited, and this is reflected in the dynamics of the
supercycles of periods $2^{N}$ via the successive formation of gaps in phase
space (the interval $-1\leq x\leq 1$) that finally give rise to the
attractor and repellor multifractal sets. To observe explicitly this process
an ensemble of initial conditions $x_{0}$ distributed uniformly across phase
space was considered and their positions were recorded at subsequent times 
\cite{robledo2, robledo3}. This set of gaps develops in time
beginning with the largest one associated with the first repellor position, then followed
by a set of two gaps associated with the next two repellor positions, next a set of four
gaps associated with the four next repellor positions, and so forth. The gaps that form
consecutively all have the same width in the logarithmic scales \cite{robledo2}, and therefore their actual widths decrease as a
power law, the same power law followed, for instance, by the position
sequence $x_{\tau }=\alpha ^{-N}$, $\tau =2^{N}$, $N=0,1,2,...$, for the
trajectory inside the attractor starting at $x_{0}=0$ (and where $\alpha \simeq 2.50291 $
is the absolute value of Feigenbaum's universal constant). The locations of
this specific family of consecutive gaps advance monotonically toward the sparsest region of the multifractal attractor located at $x=0$. See Refs. \cite{robledo1, robledo2, robledo3}.

%%%%%%%%%%%%%%%%%%%%%%%%%%%%%%%%%%%%%    SECTION 3.        %%%%%
\section{Sums of diameters as partition functions} \label{Sec3} %%%%    3. SUMS OF DIAMETERS    
The rate of convergence $W_{t}$ of an ensemble of trajectories towards any
attractor/repellor pair along the period-doubling cascade is a convenient
single-time quantity that has a straightforward definition and is practical
to implement numerically. A partition of phase space is made of $N_{b}$
equally-sized boxes or bins and a uniform distribution of $N_{c}$ initial
conditions is placed along the interval $-1\leq x\leq 1$. The ratio $N_{c}/N_{b}$ can
be adjusted to achieve optimal numerical results \cite{robledo2}. The quantity of interest 
is the number of boxes $W_{t}$ that contain trajectories  at time $t$. This rate has 
been determined for the supercycles $\overline{\mu }_{N}$, 
$N=1,2,3,...$, and its accumulation point $\mu _{\infty }$ \cite{robledo2}. 
See Fig. 19 in that reference where $W_{t}$ is shown in logarithmic scales for the first
five supercycles of periods $2^{1}$ to $2^{5}$ where we can observe the
following features: In all cases $W_{t}$ shows a similar initial and nearly
constant plateau $W_{t} \simeq \Delta $, $1\leq t\leq t_{0}$, $t_{0}=O(1)$%
, and a final well-defined decay to zero. As it can be observed in the left panel of Fig. 19
in \cite{robledo2}, the duration of the final decay grows approximately
proportionally to the period $2^{N}$ of the supercycle. There is an
intermediate slow decay of $W_{t}$ that develops as $N$ increases with
duration also just about proportional to $2^{N}$. For the shortest period $%
2^{1}$, there is no intermediate feature in $W_{t}$; this appears first for
period $2^{2}$ as a single dip and expands with one undulation every time $N$
increases by one unit. The expanding intermediate regime exhibits the
development of a power-law decay with logarithmic oscillations
(characteristic of discrete scale invariance). In the limit $N\rightarrow
\infty $ the rate takes the form $W_{t}\simeq \Delta h(\ln \tau /\ln 2)\tau
^{-B}$, $\tau =t-t_{0}$, where $h(x)$ is a periodic function with $
h(1)=1$ and $B \simeq 0.8001$ \cite{robledo2}.

The rate $W_{t}$, at the values of time for period doubling, $\tau =2^{n}$, $
n=1,2,3,...<N$, can be obtained quantitatively from the supercycle diameters 
$d_{n,m}$. Specifically,

\begin{equation}
Z_{\tau }\equiv \frac{W_{t}}{\Delta }=\sum_{m=0}^{2^{n-1}-1}d_{n,m} \label{partition1}                                       
\end{equation}

\noindent In the above expression, $ \tau =t-t_{0}=2^{n-1},\ n=1,2,3,...<N $.
Eq. (\ref{partition1}) expresses the numerical procedure followed in \cite{grassberger1}
to evaluate the exponent $B $ but it also suggests a statistical-mechanical structure if $Z_{\tau }$ is identified as a partition
function where the diameters $d_{n,m}$ play the role of configurational
terms \cite{robledo2}. The diameters $d_{N,m}$ scale with $N$ for $m$ fixed
as $d_{N,m}\simeq \alpha _{y}^{-(N-1)}$, $N$ large, where the $\alpha _{y}$
are universal constants obtained from the finite discontinuities of
Feigenbaum's trajectory scaling function $\sigma (y)=\lim_{N\rightarrow
\infty }(d_{N+1,m}/d_{N,m})$, $y=\lim_{N\rightarrow \infty }(m/2^{N})$ \cite{schuster1, robledo2}.
The largest two discontinuities of $\sigma(y)$
correspond to the sparsest and denser regions of the multifractal attractor
at $\mu _{\infty }$, for which we have, respectively, $d_{N,0}\simeq \alpha
^{-(N-1)}$ and $d_{N,1}\simeq \alpha ^{-2(N-1)}$ ($d_{1,0}=1$).

%%%%%%%%%%%%%%%%%%%%%%%%%%%%%%%%%%%%%%%%%%%%%%%%%%%%%%%%%%%%%%%%%%%%%%%%%%%%%%%%%%						%%%%% 				SECTION 4. 					%%%%%

\section{A closer analysis of the partition functions for the supercycles} \label{Sec4}          

 %%%%%%%%%%%%%%   4.  A CLOSER ANALYSIS ... %%%%%%%%%

We proceed now to study in more detail the diameters $d_{N,m}$ so that we
can evaluate the soundness of their association with configurational terms
in a partition function. With this in mind we determined their values for
the supercycles of periods $2^{N}$ from $N=1$ to $N=12$, that is, starting
with the case of a single diameter $d_{1,0}=1$ and following successively up
to a set of $2048$ diameters $d_{12,m}$, $m=0,1,...,2^{12}-1$. This task                
required the precise evaluation of the control parameter values $\overline{      
\mu }_{N}$, $N=1,...,12$.

\begin{figure}[h]%[htbp]
\centering
\epsfig{file=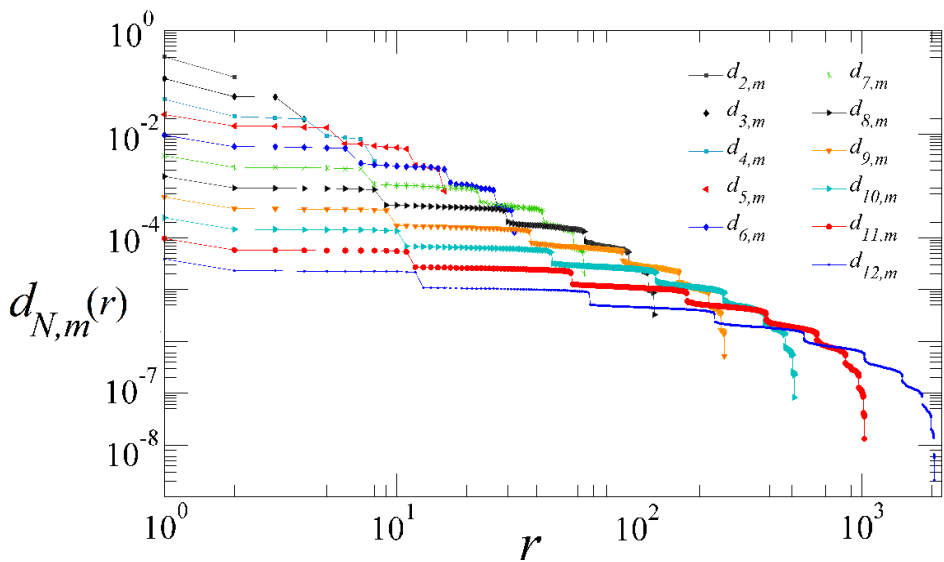, width=.8\textwidth} 
\caption{\footnotesize Length-rank distributions of diameters $d_{N,m}$ for the first $12$
supercycles in logarithmic scales. The distributions have a downhill 
terraced or multiple-plateau structure and the diameters form well-defined  
length groups. Their values for fixed period $2^{N}$ decrease on average as
$\log_{10}\alpha \simeq 0.39844$ from group to group with $N$ fixed or from $N$ to $N + 1$ for the 
same kind of group.}
\label{alldiams}
\end{figure}

In Fig. \ref{alldiams} we show the  lengths of these sets when arranged with decreasing values, 
namely, we present the $d_{N,m}$ as a function of their rank $r$, the size-rank distributions, in 
logarithmic scales, as it is often done for these type of distributions that exhibit frequently
power law behavior. We observe in Fig. \ref{alldiams} that the
distributions have a downhill terraced (or multiple-plateau) structure, the
diameters form well-defined size groups and these sizes decrease on average a fixed amount (in the logarithmic scales shown) equal to $\log_{10}\alpha \simeq 0.39844$ from group to group. This amount  reflects the well-known \cite{schuster1, hilborn1} power-law scaling of diameter sizes via the universal constant $\alpha$. Their size-rank distributions satisfy a piecewise Zipf-like law.                                                       
For example, for the largest diameter we have $d_{N,0}/d_{N+1,0}\simeq \alpha ^{1}$, whereas for
the smallest we have $d_{N,1}/d_{N+1,1}\simeq \alpha ^{2}$. Therefore $d_{N,0}\simeq \alpha ^{-N}$ and $d_{N,1}\simeq \alpha ^{-2N}$. 

\begin{figure}[h]
\centering
\epsfig{file=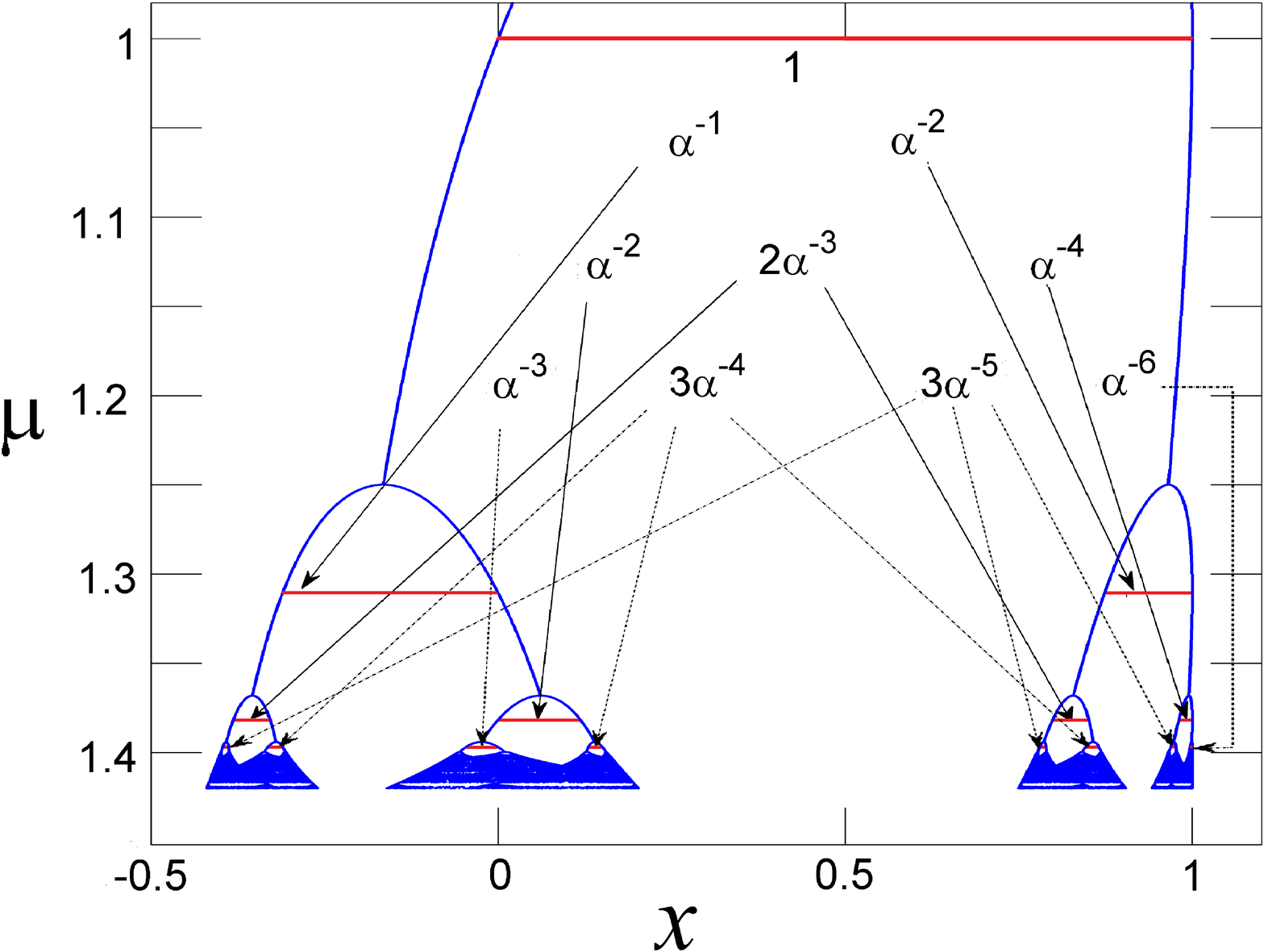, width=.8\textwidth} 
\caption{\footnotesize Sector of the bifurcation tree for the logistic map $f_{\mu }(x)$      
that shows the formation of a Pascal Triangle of diameter lengths according              
to the binomial approximation explained in the text, where $\alpha \simeq $ $%
2.50291$ is the absolute value of Fiegenbaum's universal constant.}
\label{pascalt}
\end{figure}

We observe clearly in Figs. \ref{alldiams} and \ref{N12diams} that the diameter lengths
within each group are not equal, so that there 
is no degeneracy in them. However the differences in lengths within groups diminishes 
rapidly as $N$ increases. There are two groups with only one member, the largest
and the shortest diameters, and the numbers within each group grow
monotonically from each end towards the middle-sized length group. The
numbers of diameters forming these groups can be neatly arranged into a
Pascal Triangle (see Fig. \ref{pascalt}), and therefore we anticipate the action of the
Central Limit Theorem, in a form reminiscent of the De Moivre-Laplace
theorem, so that in the limit $N\rightarrow \infty $, the
middle-sized-length group of diameters dominates the partition function $
Z_{\tau }$ and a situation similar to the saddle-point approximation occurs.
Also, in the limit $N\rightarrow \infty $ the lengths of the dominant group
(as well as those of all other groups of diameters with smaller lengths)
become closer in size (see the trend in Fig. \ref{alldiams}), so that in the limit $
N\rightarrow \infty $ there appears a true degeneracy in the dominant
partition function configurations that gives the statistical-mechanical
structure the required characteristics for ensemble equivalence and the
Legendre transform property central to statistical mechanics.

The above facts and understanding allow us to be more precise and we denote now the diameters as $d_{N,l,i}$ where the subindexes $l$ and $i$ provide more specific information than
the former subindex $m$. Subindex $l=0,1,...,N-1$ designates the group terrace (as in Fig. \ref{alldiams}), with decreasing size as $l$ increases, and the subindex $i=1,2,...,\binom{N-1}{l}$ identifies the individual diameter within the group, again with decreasing size as $i$ increases. The diameters $d_{N,l,i}$ are written as $d_{N,l,i} = A_{N,l,i}\alpha^{-(N-1-l)} \alpha^{-2l}$ where the scaling factors $\alpha ^{-(N-1-l)}$ and $\alpha^{-2l}$ give the size of the group terrace (see Fig. \ref{pascalt}) and the amplitude $A_{N,l,i}$ fixes the value of the individual diameter. (Due to the De Moivre-Laplace theorem, in the limit $N \rightarrow \infty$ the amplitude $A_{N,N/2,i}$ can be obtained from an inverse complementary error function, but we do not expand on this here). The scaling factor $\alpha^{-(N-1+l)}=\alpha ^{-(N-1-l)}\alpha^{-2l}$ can be rewritten exactly as a $q$-exponential, defined as $\exp_{q}(x)\equiv [1+(1-q)x]^{1/(1-q)}$, via use of the identity $\alpha^{-(N-1+l)}\equiv(1+\epsilon_{l})^{-\ln\alpha/\ln2}$, $\epsilon _{l} = 2^{N-1+l}-1$. That is, $d_{N,l,i} \simeq A_{N,l,i}\exp_{q}(-\beta \epsilon_{l})$, where, $q=1+\beta^{-1}$, $\beta=\ln \alpha/\ln 2$. Similarly, $Z_{\tau}\simeq \tau ^{-B }$ can be expressed as $Z_{\tau }\simeq \exp_{Q}(- B \epsilon )$, where $Q=1+B ^{-1}$ and $\epsilon =\tau -1=2^{N-1}-1$. Therefore, taking the above into account in Eq. (\ref{partition1}) we have

\begin{equation}
\exp _{Q}(-B \epsilon) \simeq \sum _{l,i} A_{N,l,i} \exp _{q}(-\beta \epsilon_{l}),             
\label{partition2} 
\end{equation}%

\noindent Eq. (\ref{partition2}) resembles a basic statistical-mechanical expression
except for the presence of the amplitudes $A_{N,l,i}$ and the fact that $q$-deformed exponential weights appear in place of ordinary exponential weights (that are recovered when $Q=q=1$). 

To explore further we use a binomial approximation for $Z_{\tau }$ 
\cite{robledo2}. That is, we adopt the approximation of considering the diameter
lengths in each group to actually have equal length ($A_{N,l,i}=1$) and assume that this
common lengths are given by the binomial combination of the scale factors of
those diameters that converge to the most crowded and most sparse regions of
the multifractal attractor. Namely, the $2^{N-1}$ diameters at the $N$-th
supercycle have lengths equal to $\alpha ^{-(N-1-l)}\alpha $ $^{-2l}$ and
occur with multiplicities $\binom{N-1}{l}$ where $l=0,1,...,N-1$. See Fig. \ref{pascalt}. The imposed degeneracy within
groups in the diameter lengths complete the Pascal Triangle structure across
the bifurcation cascade. This feature significantly simplifies the
evaluation of the partition function in Eq. (\ref{partition1}) and directly
yields
\begin{equation}
Z_{\tau }=\sum_{l=0}^{N-1} \binom{N-1}{l} \alpha ^{-(N-1-l)}\alpha ^{-2l}=\left( \alpha ^{-1}+\alpha
^{-2}\right) ^{N-1} \label{partition3}
\end{equation}

\noindent where $ \tau =2^{N-1}$. We obtain $B =0.8386$, and $Q=2.1924$, a surprisingly           
good approximation when compared to the numerical estimates $B =0.8001$
and $Q=2.2498$ of the exact values \cite{robledo2}. Eq. (\ref{partition2})
reads now
\begin{equation}
\exp _{Q}(-\beta F )=\sum_{l=0}^{N-1}\Omega (N-1,l)\exp _{q}(-\beta
\epsilon _l),  \label{partition4}
\end{equation}%
where $F/\epsilon = (1-q)/(1-Q)$, and $\Omega (N-1,l)= \binom{N-1}{l}$, $\alpha ^{-(N-1-l)}\alpha ^{-2l} = 2^{-(N-1+l)(\ln \alpha /\ln 2)} = \exp_{q}(-\beta \epsilon _l)$.
In the language of thermal systems Eq. (\ref{partition4}) reads as follows: There are $N-1$ degrees of freedom that generate $2^{N-1}$ configurations, and these occupy $N$ energy levels with 
degeneracies $\Omega (N-1,l)$, $l=0,1,...,N-1$. Under the binomial approximation the energies of the $2^{N-1}$ configurations become confined into the energy values $\epsilon_{l}=2^{(N-1+l)}$, $l=0,1,...,N-1$. In the generalized canonical partition function all the $q$-exponential weights acquire a fixed inverse temperature $\beta=\ln \alpha /\ln 2$. When we extend the study of the quadratic map to the infinite family of unimodal maps with extremum of nonlinearity $1<z<\infty$ the inverse temperature $\beta$ can be varied continuously, as the universal constant $\alpha(z)$ varies monotonically with $z$ \cite{capel1}. 

%%%%%%%%%%%                             SECTION 5. 						%%%

\section{A limiting statistical-mechanical structure for the dynamics at the Feigenbaum point} \label{Sec5}      

          %%%%%%%%%%%   5. A LIMITING STATISTICAL-MECHANICAL STRUCURE %%%%

\begin{figure}[h]
\centering
\epsfig{file=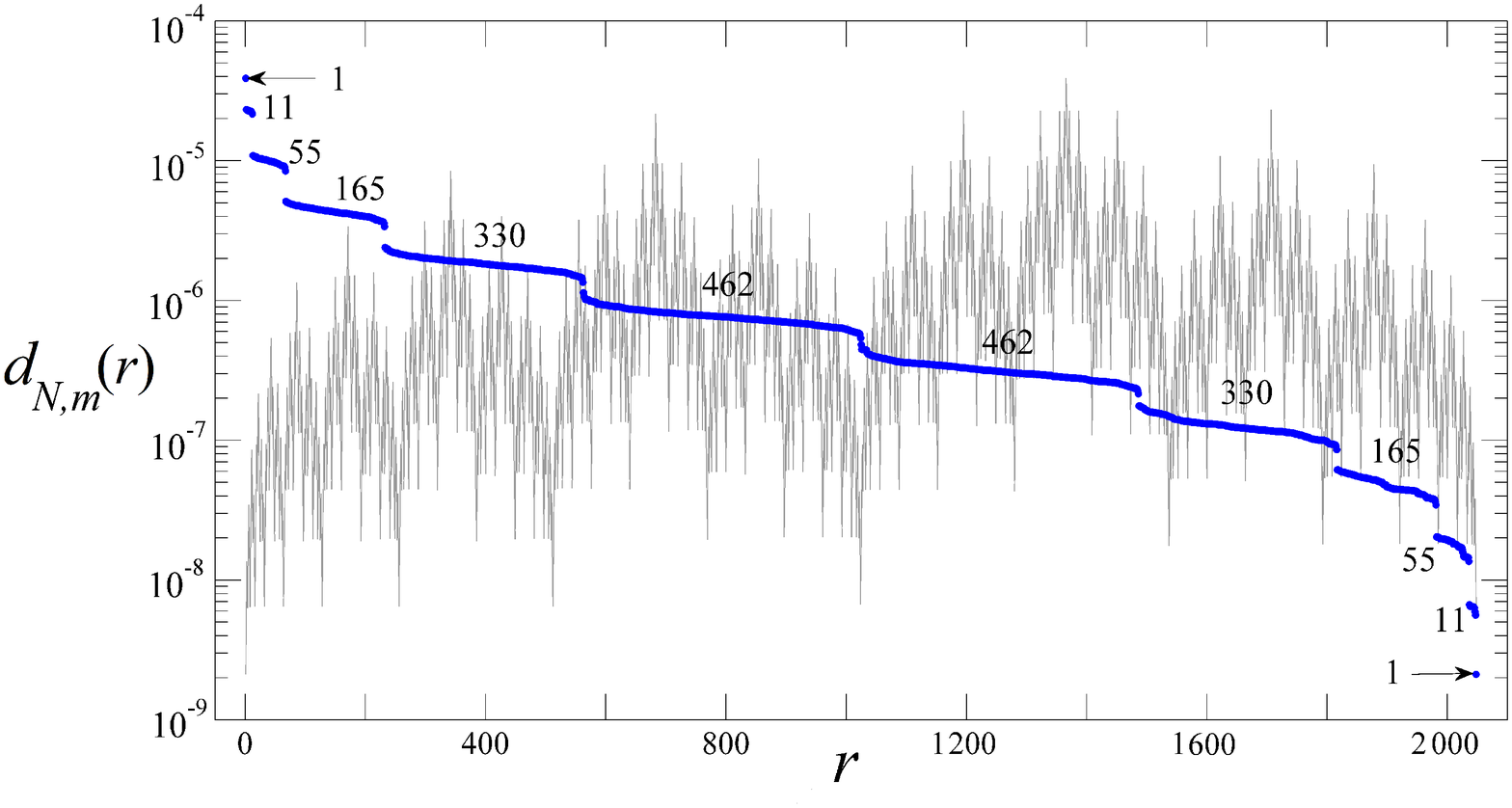, width=.8\textwidth}
\caption{\footnotesize Length-rank distribution for the $2048$ diameters of the $12$th                 
supercycle in semi-logarithmic scales. The figures indicate the number of                  
diameters in each group. As $N$ grows the length-rank distributions approach
the binomial size-rank distribution and the De Moivre-Laplace theorem
applies at the transition to chaos. See text. In the background we show the same diameters 
before sorting them out as a function of $m$ }
\label{N12diams}
\end{figure}

According to our scheme, for finite $N$ (the supercycle of period $2^{N}$ at 
$\overline{\mu }_{N}$) we can form $N-1$ partition functions $Z_{\tau }$, $%
\tau =2^{n-1}$,$\ n=1,2,3,...,N-1$. The number of terms in these partition
functions range from a single term, $d_{1,0}$, to $2^{N-1}$ terms, $d_{N,m}$%
, $m=0,1,2,...,2^{N-1}-1$. As explained, for uniform distributions of
initial conditions $-1\leq x_{0}\leq 1$ at $\mu =\overline{\mu }_{N}$, the
partition functions $Z_{\tau }$ measure the fraction of ensemble
trajectories still away from the attractor at times $\tau =2^{n-1}$,$\
n=1,2,3,...,N-1$. These times coincide with the sequential process of
phase-space gap formation by the trajectories \cite{robledo2}. The gaps correspond to the intervals in $-1\leq x\leq 1$ located between the bifurcation forks in the period-doubling cascade, when $%
\mu =\overline{\mu }_{N}$, that is, the gap intervals are placed between
consecutive diameters. As $N$\ grows new smaller gaps proliferate while the
new diameters grow in number and each of them decreases in value. See Fig. \ref{N12diams}
where the numbers of diameters are shown for each group formed for the case of the 
12th supercycle. The number of groups into which the diameters distribute increases 
as it does the number of diameters within each group. As we have indicated these 
increments obey the entries in the Pascal Triangle generated by a binomial. Although
the diameters within each group are never equal their differences decrease
rapidly. The dominant term in $Z_{\tau }$ is that associated with $%
\Omega (N-1,(N-1)/2)$, $N$ odd, and\ in the limit $N\rightarrow \infty $ we have that $%
Z_{\infty }=$ $\Omega (N\rightarrow \infty , \ l=N/2\rightarrow \infty )$. We
interpret this last equality as ensemble equivalence in the thermodynamic
limit (here $N\rightarrow \infty $ is the attractor at the transition to
chaos).

It is more convenient to describe the ensemble equivalence in terms of the
binomial approximation of the partition function $Z_{\tau }$ given by Eqs. (\ref{partition3}) and (\ref{partition4}), where $\Omega (N-1,l)$ plays the
role of a `microcanonical' partition function representing the system
configurations with fixed diameter length $\alpha
^{-(N-1-l)}\alpha $ $^{-2l}$ and $Z_{\tau }$ stands for the `canonical'
partition function that is formed by weighting the degenerate configurations 
$\Omega (N-1,l)$ for each length group by the factor $\alpha ^{-(N-1+l)}\equiv
\exp _{q}(-\beta \epsilon_{l})$. According to the De Moivre-Laplace early form of
the Central Limit Theorem the growth of $N$ drives the binomial distribution
towards a Gaussian distribution
\begin{equation}
\delta ^{N}P_{l,\ N-l}\simeq \frac{1}{\sqrt{2\pi N\rho \sigma }}\exp \left( -%
\frac{x^{2}}{2N\rho \sigma }\right) ,  \label{demoivrelaplace1}
\end{equation}%
where $\delta =\alpha ^{-1}+\alpha ^{-2}$, $\rho \sim \alpha ^{-2}$, $\sigma
\sim \alpha ^{-1}$ and $x=l-N\rho$. For large $N$ the
midpoint terms in the expansion of the binomial dominate, $Z_{\tau }=$ $%
\left( \alpha ^{-1}+\alpha ^{-2}\right) ^{N-1}\simeq $ $\binom{N}{N/2} \alpha ^{-3N/2}\sim 2^{N}\alpha ^{-3N/2}$, and in the limit $%
N\rightarrow \infty $ we have that $Z_{\infty }=$ $\Omega (N\rightarrow
\infty , \ l=N/2\rightarrow \infty )$. For $N$ fixed the `energies' $\epsilon_l$ range from $2^{N-1} - 1$ to $2^{2(N-1)}-1$. For large $N$ the `energy' that corresponds to the `microcanonical' partition function that becomes the dominant term in $Z_\tau$ is $\epsilon_{N/2} \simeq 2^{3/2 N}$.

A crossover to ordinary BG type statistics takes place when $\mu \gtrsim \mu
_{\infty }$ and the attractor becomes chaotic. For $\Delta \mu \equiv \mu
-\mu _{\infty }>0$ the attractors are made up of $2^{N}$, $N=1,2,3,...$,
bands, with $N$ larger for $\Delta \mu $ smaller, while the Lyapunov exponent
scales as $\lambda \sim 2^{-N}$. The trajectories consist of an interband
periodic motion of period $2^{N}$ and an intraband chaotic motion. As explained in Ref. \cite{robledo2} the consideration of backward iterations in unimodal maps, together with the expansion of separation of trajectories when $\lambda > 0$, can be invoked to write a partition function similar to that in Eq. \ref{partition2} but now with ordinary exponentials as configurational weights.

As it is well known \cite{beck1}, the so-called thermodynamic formalism for
the description of the geometric properties of multifractal sets is built
around a statistical-mechanical framework of the BG type. The partition
function formulated to study multifractal properties, like the spectrum of
singularities $f(\widetilde{\alpha })$, is written as%
\begin{equation}
Z(\widetilde{\tau },\widetilde{\beta })=\sum_{m}^{M}p_{m}^{\widetilde{\tau }%
}l_{m}^{\widetilde{\beta }},  \label{partition6}
\end{equation}%
where the $l_{m}$ in one-dimensional systems are $M$ disjoint interval
lengths that cover the multifractal set and the $p_{m}$ are probabilities
given to these intervals. The standard practice consists of demanding that $%
Z(\widetilde{\tau },\widetilde{\beta })$ neither vanishes nor diverges in
the limit $l_{m}\rightarrow 0$ for all $m$ (notice that in this limit $M\rightarrow \infty 
$) . Under this condition the exponents $\widetilde{\tau }$ and $\widetilde{%
\beta }$ define a function $\widetilde{\tau }(\widetilde{\beta })$ from
which $f(\widetilde{\alpha })$ is obtained via Legendre transformation \cite%
{beck1}. When the multifractal is an attractor its elements are ordered
dynamically, and for the Feigenbaum attractor the trajectory with initial
condition $x_{0}=0$ generates in succession the positions that form the
diameters, generating the entire set of diameters $d_{N,m}$, $m = 0,1,2,...,2^{(N-1)}-1$,
$N=1,2,...$. 
Because the diameters cover the
attractor it is natural to choose the covering lengths at stage $N$ to be $%
l_{m}^{(N)}=d_{N,m}$ and to assign to each of them the same probability $%
p_{m}^{(N)}=\left( 1/2\right) ^{N-1}$, and the condition $Z(\widetilde{\tau }%
,\widetilde{\beta })=1$ reproduces Eq. (\ref{partition1}) when $%
p_{m}^{(N)}=\tau ^{-1}=\left( 1/2\right) ^{N-1}$ with $\widetilde{\tau }%
=-B $ and $\widetilde{\beta }=1$.

%%%%%%%%%%%%%%%%%%%%%%%%%%%%%%%%%%%%%%%%%%%%%%%%%%%%%%%%%%%%%%%%%%%%%%%%%%%											%%          SECTION 6.      %%

\section{Summary and discussion}  \label{Sec6}                      																 %%%%%%%%%%%%%  6. SUMMARY AND DISCUSSION  %%%%%%%%%%%%%%%%%%%

 The items we studied are the following: i) The partition function we considered 
is the sum of attractor position distances (the so-called diameters of the
supercycles \cite{schuster1, hilborn1}) for each period $2^N$ along the
bifurcation cascade that leads to the transition to chaos. ii) For
uniformly-distributed sets of initial conditions $x_{0}$ the partition
function is equal to the number of bins that still contain trajectories en
route to the attractor at time $\tau =2^{n}$, $n=1,2,3,... \lesssim N$, where the
supercycle period is $\tau =2^{N}$, $N>1$. iii) For $N$ fixed the values of
the diameters distribute into well-defined groups with a size-rank structure
that develops into a power law as $N$ increases. These groups can be
arranged into a Pascal Triangle when considering all $N$ up to $N \rightarrow \infty$,  
but the diameters within each group are not equal. Nevertheless, their
differences diminish rapidly as $N$ increases, so that a binomial
\textit{approximation} can be introduced such that the diameters within each group
are considered equal for all $N$. iv) In the limit $N \rightarrow \infty $
the diameter-group degeneracy imparts the partition function the required
structure to observe ensemble equivalence, and other familiar features of
statistical mechanics, even though the configurational weights are not
exponential. v) The visible or `macroscopic'\ manifestation of the
statistical-mechanical structure, the emergence of a power law with
log-periodic modulation associated with the rate of approach of trajectories
towards the Feigenbaum attractor, is linked to the sequential process of
phase-space gap formation. vi) Beyond the transition to chaos, when the
attractors become sets of chaotic bands, the configurational weights are
converted into ordinary exponentials and the usual BG form is recovered. 

The main advance presented here with respect to Ref. \cite{robledo2} is the determination of the terrace structure displayed by the diameters for finite $N$ shown in Figs. \ref{alldiams} and \ref{N12diams}. This fact allowed us to write the partition function in Eq. \ref{partition1} explicitly as Eq. \ref{partition2}. Therefore we were able to study how the lack of configurational degeneracy gradually disappears as $N \rightarrow \infty$ leading to ensemble equivalence.

Chaotic dynamics in nonlinear systems accepts statistical-mechanical
descriptions \cite{beck1}. Unimodal maps, usually represented by the
logistic map, offer a simple but nontrivial model system in which to explore
the development of such a statistical-mechanical structure, to examine the
gradual fulfilment of basic elements and eventually the display of the full
ordinary features of the BG formalism. A unimodal map is a well-defined and
controllable numerical laboratory for the observation of the limit of
validity of the BG formalism when the ergodic and mixing properties of
chaotic dynamics break down. As it as long been known unimodal maps display two bifurcation cascades
that take place in opposite directions in control parameter space, one for $%
\mu <\mu _{\infty }$ when periodic attractors double their periods, and the
other for $\mu >\mu _{\infty }$ when chaotic-band attractors attractors
split doubling their number of bands. The two cascades meet at $\mu = \mu
_{\infty }$. Infinitely many reproductions of these inverse cascades appear
within the windows of periodicity that interrupt the chaotic-band attractors
for $\mu >\mu _{\infty }$ \cite{schuster1, hilborn1}.

As we have mentioned, the ergodic and mixing trajectories of chaotic-band
attractors conform to a statistical mechanical-structure of the BG type \cite%
{beck1}. We have described that the positions of
periodic attractors can be used to define partition functions and that these
capture information on the dynamics towards the attractors \cite{robledo2}.
However, as we have explained, these partition functions lack some standard
properties required in a thermodynamic formalism, such as the degeneracy of
configurational states that manifests as ensemble equivalence and the
correspondence of their respective thermodynamic potentials in the
thermodynamic limit (that in the unimodal map model is the limit $%
N\rightarrow \infty $ of infinite period). For finite $N$ the
configurational terms (diameters $d_{N,m}$) separate into well-defined
magnitude (length) groups but they are not equal within each group. These
groups of diameters are the prototypes of `microcanonical' ensembles while
the consideration of all groups, all diameters for a given supercycle of period $2^N$ is
the candidate version of the `canonical' ensemble. As we have seen, when $%
N\rightarrow \infty $ the diameters seem to fulfill a binomial approximation such
that the (vanishing) lengths within the dominant diameter groups (with divergent numbers) 
become equal and the De Moivre-Laplace theorem establishes the equivalence between the
`microcanonical' and `canonical' ensembles.
The binomial approximation we presented for finite $N$ allows for a conventional interpretation in the language of thermal systems.

\section*{Acknowledgments}
Support by DGAPA-UNAM-IN100311 and CONACyT-CB-2011-167978 (Mexican Agencies) is acknowledged.

\end{document}